\documentstyle[emulateapj]{article}

\def\ltsima{$\; \buildrel < \over \sim \;$}
\def\simlt{\lower.5ex\hbox{\ltsima}}
\def\gtsima{$\; \buildrel > \over \sim \;$}
\def\simgt{\lower.5ex\hbox{\gtsima}}
\def\be{\begin{equation}}
\def\ee{\end{equation}}
\def\kms{{\rm\,km\,s^{-1}}}

\begin{document}
\title{Possible Evidence for a Truncated Thin Disk in Seyfert 1 Galaxy NGC4593} 
\author{ Youjun Lu\altaffilmark{1,2,3,4} and Tinggui Wang\altaffilmark{1,2,4}}
\altaffiltext{1}{Center for Astrophysics, Univ. of Sci. \& Tech. of China, Hefei,
Anhui 230026, P.R. China}
\altaffiltext{2}{National Astronomical Observatories, Chinese Academy of Sciences }
\altaffiltext{3}{Current Address: Princeton University Observatory, Princeton,
NJ 08544}
\altaffiltext{4}{Email: lyj@astro.princeton.edu (YL); twang@ustc.edu.cn (TW)}

\begin{abstract}
We compile the multi-band spectral energy distribution (SED) of Seyfert
1 galaxy NGC4593, and investigate its optical-to-ultraviolet (OUV) 
continuum and iron K$\alpha$ line profile. 
No strong ``big blue bump'' is found in OUV band.
The iron K$\alpha$ line profile is not as broad as expected
from a disk which extends down to the last stable orbit. Both of them
can be modeled by a truncated thin disk, and support the existence
of a truncated thin disk in the system. It is of much interest
that NGC4593 may be a typical object with an accretion rate $\dot{m}=\dot{M}
/M_{Edd} \sim
0.01-0.1$ and harbors a two component accretion flow: an outer thin disk and
an inner hot flow--possibly ADAF, analogous to the low and intermediate
state of low mass X-ray binaries (LMXB).
\end{abstract}


 
\section{Introduction}

It is currently known that accretion onto black holes can take four forms:
the standard thin disk (\cite{ss}); optically thin Advection-Dominated
Accretion Flows (ADAFs: \cite{ich}, Rees et al. 1982; Narayan \& Yi 1994,
1995a, 1995b; Abramowicz et al. 1995); optically thick ADAFs (Abramowicz
et al. 1998); and geometrically thin disk with small
advective and larger viscosity (\cite{sle}). The last form is thermally
unstable (\cite{piran}), and therefore is unlikely to be
considered viable for real flows. The other three forms are all thought to
exist in Active Galactic Nuclei (AGN). Evidence has been shown that some
Seyfert galaxies and QSOs may accrete material via thin disk (\cite{kriss}),
while Low Luminosity AGNs (LLAGNs) may accrete material via ADAFs at
very low accretion rates and their continuum emission can be well
explained by ADAFs (\cite{dima}, \cite{qn}).

Lasota et al. (1996) and Narayan (1996) proposed
that in some luminous AGNs accretion may occur in an intermediate form,
i.e. that the accretion flow consists of an outer standard disk and an
inner ADAF in between a transition radius $r_{tr}$.
$r_{tr}$ may be directly determined by the accretion rate: the higher
accretion rate, the smaller $r_{tr}$ until it reaches the last stable orbit.
This idea has been successfully used to model the spectra of the
soft- and hard-state of some LMXBs,
which are thought to be a small scale version of AGNs (\cite{nbm};
\cite{emn}). Recent research on NGC4258, M81 and NGC4579 also
demonstrate the existence of this kind of two component flows in LLAGNs
(\cite{gnb}; \cite{quat}). However, there are few examples for such an
accretion configuration in luminous Seyfert 1 galaxies, although the
distribution of the X-ray spectral index against the accretion rate
suggests that this two component flow may be common in Seyfert galaxies
(\cite{luyu}).

NGC4593 is a normal Seyfert 1 galaxy with redshift $z=0.009$. The mass of
its central black hole is determined to be $1.4\times 10^7{M}_{\odot}$
(\cite{kd}) using the reverberation mapping method, which is believed to
give a reliable mass estimation (\cite{pw}).
The mass is also constrained to be $\simlt 2.3\times 10^7{M}_{\odot}$
if the observed redshift of the central component of H$\alpha$
to the narrow line component is due to gravitational redshift (\cite{kd}).
We adopt the mass as $1.4\times 10^7{M}_{\odot}$,
and then the Eddington ratio is
estimated to be $\sim 0.01$. In addition, NGC4593 is one of the objects
which have been suggested to accrete material via an outer thin disk and an
inner ADAF (\cite{luyu}). It is therefore very interesting to look in
different bands for evidence of the existence of a truncated thin disk
in NGC4593.

In this letter, we compile the nuclear continuous emission of NGC4593,
and investigate the OUV continuum and the iron K$\alpha$ line profile.
Evidence is shown that this object seems to be analogous
to the object NGC4258 which accretes material via an outer thin disk and an
inner ADAF.

\section{The nuclear continuum emission}

\subsection{Data Compilation}

At radio band,
NGC4593 has been observed by VLA at 3.6cm, 6cm and 20cm. The
radio emission comes from unresolved regions with extent less than
37pc at 6cm and 20cm (\cite{uw}), and 17pc at 3.6cm (\cite{kinney}).
These can be used as upper limits to the nuclear emission since 37pc
and 17pc are still too large to distinguish whether the emission
comes from a disk or jet. 

The measurements have been obtained using IRAS in the far infrared band.
However, IRAS has very poor spatial resolution with a beam $\simgt 1\arcmin$.
The measurements can only provide a poorly constrained upper limit to the
nuclear emission because the galaxy contributes at least 50\%
to the measurements (Santos-Lle\'{o} et al. 1995). 

In the ultraviolet-to-optical and J-, H-, K- and L-bands, 
Santos-Lle\'{o} et al. (1994, 1995)
have obtained the nuclear continuum emission within a campaign 
to monitor spectral variations of NGC4593. They modeled and 
subtracted the contribution of the stellar population to the
nuclear continuous emission, and also corrected for Galactic
extinction. The ROSAT spectrum shows that there is no significant intrinsic
extinction in NGC4593 (\cite{wlz}).
The absolute UV fluxes are difficult to define owing to the
flux variability. However, the relative spectral shape didn't
change much based on Santos-Lle\'{o} et al. (1994, 1995) data.
We adopt the mean flux in this letter.

NGC4593 was observed by ROSAT at soft X-ray band. The spectrum can
be fitted by a power-law with spectral index $1.08\pm0.22$
(\cite{wlz}). In the hard X-ray, ASCA observed NGC4593 three
times; we use the longest observation taken in 1994 in this letter.
The continuum between 1.5-4.5keV and 7.2-10keV can be modeled by
an absorbed power-law with spectral index $0.86\pm0.03$
(see the data processing procedure in section 3).

We adopt ${H_0=75\kms\rm Mpc^{-1}}$, and then calculate
the monochromatic luminosities. The SED are shown in Fig~1. There is
a discontinuity between {\it ROSAT} and {\it ASCA} band which should be
caused by the X-ray variability. Previous observations by {\it EXOSAT}
have been revealed that the X-ray flux at 2-10keV band varies in a range of
$(4.9-1.1)\times10^{-11}$ cgs (\cite{san94}), hence NGC4593 was in a ``high
state'' when it was observed by {\it ASCA} (with 2-10keV flux of
$3.5\times10^{-11}$ cgs), while it was in a ``low state'' when {\it ROSAT}
observed it (with an extrapolated 2-10keV flux of $\sim 10^{-11}$cgs).
The SED of NGC4258, which is compiled by Chary et al. (2000), is also shown
in Fig~1. It is surprising that NGC4593 has an almost identical infrared
to hard X-ray SED shape (at the ``low state'') with that of NGC4258,
although its luminosity is larger
than that of NGC4258 by a factor of about 30. The multi-band spectra
of NGC4258 is well explained by a two component accretion flow model
with a transition radius at 10$-$100$r_g$ (\cite{las}; \cite{gnb}).
Therefore, the radio-to-X-ray spectra might also be explained by such a model.

\subsection{Truncated thin disk model for the OUV spectrum}

The emission from a thin accretion disk can be roughly modeled as
a multi-color black body (e.g. \cite{fkr}). It is independent
of the microphysics of the disk and depends only on the mass of
black hole $M$, the accretion rate $\dot{M}=\dot{m}\dot{M}_{Edd}$
($\dot{M}_{Edd}=10L_{Edd}/c^2$ is the Eddington accretion rate, $\dot{m}$
is a dimensionless accretion rate)
and the inner edge of the disk $r_{in}$ if we specify the outer
edge as $r_{out}=10^5r_g$ ($r_g=2GM/c^2$ is the gravitational
radius) of the disk and the inclination of the disk to the
line of sight as $i=30\arcdeg$. 

We first model the OUV continuum spectrum as emission from
a standard thin disk extending down to the last stable orbit ($r_{in}=3r_{\rm g}$).
The dotted lines in Fig~2 show the standard thin disk models with (from
top to bottom) $\dot{m}=$0.008, 0.02, 0.06 and 0.20.
None of these models is capable of explaining the OUV
nuclear continuum emission. Then, we model the OUV spectrum as
emission from a truncated thin disk. The solid line in Fig~2 presents
the emission from a truncated thin disk model by setting $r_{in}\approx 30r_g$
and $\dot{m}\approx 0.055$. This model can fit the
OUV nuclear continuum emission well. Though it does not produce
enough energy emission at the J-, H-, L- and K-band, this is not a serious
difficulty because most of the near infrared emission is possibly emitted in
an outer hot dust torus, which is illuminated by the inner UV-X-ray source
(with a total energy of $\sim 10^{43}$cgs in UV-X-ray band; Santos-Lle\'{o}
et al. 1995) but not the thin disk. In addition, the heating of the outer
thin disk by the illumination of the central X-ray emission will change the
OUV emission, but not much due to the small subtending angle
of the central X-ray emission.
If the above interpretation of the OUV nuclear continuum 
emission of NGC4593 is correct, this should be a good evidence for a
a truncated disk in the nucleus.
Further observations and detail theoretical modeling of the whole spectra
are needed.

\section{Fe k$\alpha$ line }

One distinct feature of a system with a truncated outer thin disk 
and an inner ADAF is that there should not be a large relativistically
broadened iron K$\alpha$ line in the hard X-ray band.
Indeed, Nandra et al. (1997) presented a spectral fit with a Gaussian 
model, and found that the iron K$\alpha$ line is only moderately broad
in NGC4593.  The narrowness of the line is consistent
with an origin in a truncated outer thin disk and an inner ADAF but not
a thin disk extending down to the last stable orbit. In order to test
this and determine the truncated radius from the X-ray emission, we have
re-analyzed the data set of the longest observation taken in 1994.

The X-ray data have been processed with standard screening criterion. For
SIS data, the faint mode data were converted to Bright2 mode data and
corrected for dark frame error and echo effects. Hot flickering pixels
were removed from the SIS and rise-time rejection was applied to exclude
particle events from the GIS data. SIS grade 0, 2, 3, 4 data were selected
for the analysis.  The source counts were extracted from circular regions
of radius $\sim$3.5\arcmin~ and $\sim$6\arcmin~ for the SIS and the GIS,
respectively, while the background counts were extracted from a region near
the source for the GIS and from the uncontaminated region of the same CCD
chip for the SIS.
For the SIS spectrum, the response matrices appropriate for the date of
the observation 
were made using the script {\em sisrmg}. For the GIS
spectrum, the 1994 response matrices
were adopted. Ancillary response files were made for each
detector using {\em ascaarf}. The ASCA data preparation and the spectral
analysis were performed using version 1.4 of the XSELECT package and version
10.01 of XSPEC.

Since we are only interested in the iron K$\alpha$ line, only the portion
of the spectrum
between 1.5 to 10 keV are fitted to minimized the complex due to the OVII
and OVIII absorption edge in the warm absorber.  The continuum between
1.5-4.5keV and 7.2-10 keV is modeled by an absorbed power-law. This fit
yields a photon index $\Gamma=1.86\pm0.03 $, which is consistent with that
obtained by Nandra et al. (1997). The photon index is then fixed at this
value for the following more complicated fit. 

Since the disk is truncated at some radius, the disk line emission
does not extend into the last stable radius. We use a disk line model
to describe this disk emission, but leave the inner radius as a free
parameter. The outer disk radius is fixed at very large radius
$10^5r_g$ since the data suggests a large radius but does not constrained
its value well. The line emissivity index ($F_l\propto r^{-q}$) is fixed
at $q=2.0$ as expected if the general disk is
illuminated by a point source above the symmetry axis. The line energy is
fixed at 6.4keV in the source rest frame,
since the disk line is expected
to arise from fluorescence. In addition to this disk component, line emission
is also expected from the transition region between the outer thin
disk and the inner ADAF, where the gas temperature
is high enough to ionize the iron to high ionization state but still
not high enough to have the iron atoms fully stripped. This transition region
might not be very thick, therefore we use a Kepler rotating ring model to
describe it. Since this component is collisionally ionized and excited,
we fixed the line energy to 6.7keV. The width of the ring is hard to
estimate, but all fits suggest that it is very small. We fixed it to be
0.2 of the outer
radius.  Choosing other value does not seriously affect the results presented
below as long as it is not too wide. We assume the ionized line set just
inside the transition zone, so the outer radius of this ring is set to
the inner edge of the above disk.
  
This line model has 4 free parameters, the inner disk radius, the
inclination and two normalizations, which is only one more free
parameter than the free width Gaussian fit. The data can be very well
fitted by this model (Fig. 3). However, the data quality
is not sufficient to determine all parameters. We therefore fit the spectrum
for different inclinations of disk.  The best fit truncated radius
is $57^{+38}_{-16}$, $80^{+46}_{-28}$ and $132^{+85}_{-38}$GM/c$^2$ for
$i$=20\arcdeg, 30\arcdeg and 50\arcdeg, respectively. All fits are better
than a Gaussian 
line model with a $\Delta \chi^{2}$=0.5, 2.2, 3.7 for $i$=20\arcdeg,
30\arcdeg, and 50\arcdeg, respectively, for the same number of free
parameters. This again gives an estimate of the truncated radius
to be $\simgt30r_g$, which is consistent with the requirement of an
outer thin disk truncated at about $30r_g$ from the shape of OUV
continuum in last section. These results give strong evidence for
the existence of an truncated outer thin disk and an inner hot
flows in the nucleus of NGC4593, though it is hard to determine the
exact truncation radius. 
                   
\section{Discussion}

In this letter, our main result is that there may exist a truncated
thin disk in NGC4593. Both the OUV
continuum spectral shape and iron K$\alpha$ line profile, two
independent observations in different bands,
consistently require an outer thin disk truncated at about
30$r_g$, instead of extending down to the last stable orbit.
This result is very interesting because NGC4593 is a normal Seyfert 1
galaxy and much more luminous than previously known objects suspected
to have a truncated thin disk.

NGC4258 is a well known object suspected to accrete material via
two component disk (\cite{las}, \cite{gnb}).
Chary et al. (2000) argued that its infrared
spectrum ($\propto \nu~^{-1.4\pm0.1}$) obviates the need for a substantial
contribution from an outer thin disk, and can be interpreted by a pure
ADAF.
The lack of nuclear emission data in OUV bands, which should give
a strong constraint on the existence of a truncated thin disk, makes it
hard to ascertain whether the accretion flow consists of two component or only a
pure ADAF. NGC4593 has a similar SED shape with that of NGC4258 and,
furthermore, the nuclear emission data in OUV bands are available.
Its OUV spectra can well be modeled by a truncated thin disk.
If the inferred black hole mass and the interpretation of
the OUV continuum are both correct, then
$\dot{m}$ is suggested to be about 0.055. Theoretically,
the accretion flow of a system with such an accretion rate
probably has the configuration of an outer thin disk and an inner ADAF,
which has been studied in detail by Esin, McClintock \& Narayan (1997)
for LMXB. If the true mass of the
central black hole is larger than the inferred one by an order of magnitude,
then it may also be possible to interpret the SED of NGC4593 by a pure ADAF.
However, the turn over at OUV band cannot be explained by such
a pure ADAF. 
It is emphasized here that NGC4593 might be a good candidate
for the implication of the two component accretion flows.

The change in the X-ray spectrum from a low state to a high state may correspond
with the variation of the truncated radius which is principally caused by the
variation of the accretion rate. The dynamics of how a transition from an
outer thin disk to an inner ADAF
in the accretion flow is still poorly understood, the only known physical
process of ``evaporation'' possibly takes the responsibility for this change
(\cite{mm}). Further model fitting of the multi-band spectra for NGC4593 
will be presented in future work, and it is of theoretical interest to constrain
the dynamics of the transition by this fitting.

The iron K$\alpha$ line in NGC4593 is not like the relativistic broadening
one seen in MCG-6-30-15 (\cite{tan}, \cite{iwa}), but is
moderately broad and requires the thin disk to be truncated at a radius
$\simgt 30r_g$. Similar results were reported in recent analyses of 
the ASCA spectrum of NGC5548 (\cite{chiang}) and of the ASCA and
RXTE spectra of IC4329A (\cite{done}),
where the lines are broad, but not as broad as
expected from a disk extending down to the last stable orbit.
Moreover, the fact that amount of relativistically smeared reflection is
rather less than unity for IC4329A (\cite{done}), 
is inconsistent with a cold disk extending down to the last stable orbit,
and also supports the existence of a truncated thin disk in the system. 
These results are consistent with the classification of these objects 
as having the configuration of an outer thin disk and an
inner ADAF (\cite{luyu}). This highlights the possibility that
the two component accretion flow may be common in Seyfert 1 galaxies.
However, using the BeppoSAX data, Guainazzi et al. (1999) reported that the
amount of reflection and iron line properties are consistent with an
illuminated cold disk down to $3r_{\rm g}$. This is inconsistent with 
our result. The incosistency is partly caused by the lower spectral
resolution of BeppoSAX (than ASCA), which causes the line profile to be
less constrained. Moreover, the line profile is also broadened by inclusion
of a sharp edge in their model, which is not a correct assumption since it
would be smoothed by relativistic effects.

It should be emphasized here that a two component model, a 6.7keV line from the
transition region and a 6.4keV fluorescence line from an outer thin disk,
can well fit the iron line in NGC4593 in this letter.
ASCA observations detected an iron line with a centroid energy $\simeq
6.7$keV in LLAGNs (see \cite{ishi} for M81 and \cite{tera} for NGC4579),
which is incompatible with the 6.4keV fluorescent
line expected from a thin disk as in the typical case of MCG-6-30-15. For
these systems (say M81 and NGC4579), the inner edge of a thin disk (at the
transition radius) lies at around 100$r_g$ (\cite{quat}), thus there is
no significant fluorescent line emission from the outer thin disk. If the
inner edge goes in (possibly with an increasing accretion rate), some
fluorescent line emission from the outer thin disk will contribute as
is the case in NGC4593. It may be also true for NGC5548 and IC4329A.
Future X-ray spectroscopy observations with {\it Chandra } and {\it 
XMM-Newton} should give a further test of the two component accretion flow. 

We summarize our main result as the existence of a truncated thin disk
in the nucleus of NGC4593. NGC4593 may be a typical candidate which
accretes material via an outer thin disk and an inner ADAF with 
accretion rate $\dot{m}\sim 0.01-0.1$. Detail investigation on those
objects 
which are suspected to accrete material via an outer thin disk and an 
inner ADAF, 
is crucial in establishing two component accretion flows on firm footing. 

\acknowledgments{ 
We thank Dr. S.P. Oh for a careful reading and an anonymous referee for
helpful comments. TW acknowledges the financial support from Chinese
National Science Foundation (NSF-19925313). YL thanks the hospitality
of the Department of Astrophysical Sciences, Princeton University.
}

\begin{figure}
\figurenum{1}
\epsscale{1.00}
\plotfiddle{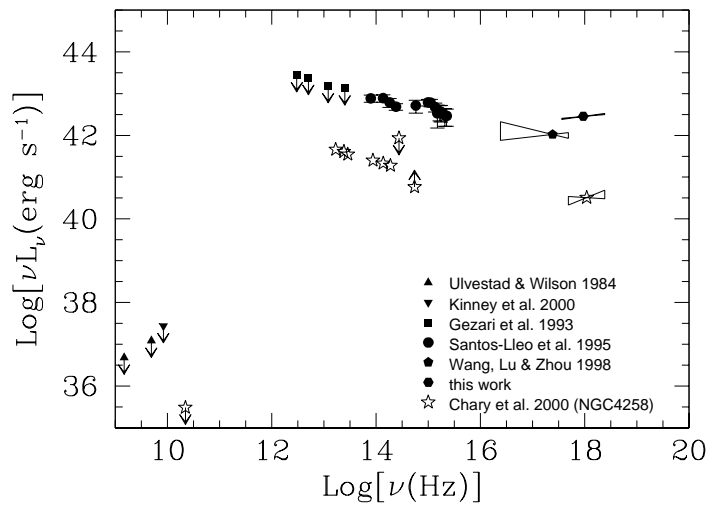}{7.cm}{270}{45}{45}{-200}{250}
\caption[lu_figure1.ps]{ Spectral Energy Distribution of the nuclei of
NGC4593 and NGC4258. The filled points and the stars
represent NGC4593 and NGC4258, respectively.  \label{fig-1}}
\end{figure}


\begin{figure}
\figurenum{2}
\epsscale{1.00}
\plotfiddle{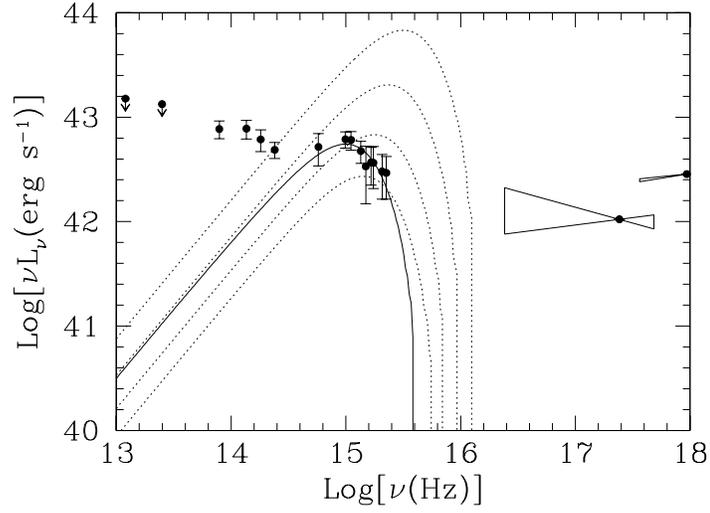}{7.cm}{270}{45}{45}{-200}{250}
\caption[lu_figure2.ps]{Multicolor black body thin accretion disk models
for the optical-to-UV emission of NGC4593: Dotted lines from top to
bottom: $\dot{m}$=0.2, 0.06, 0.02 and 0.008; solid line: $\dot{m}=0.055$
and $r_{tr}=30r_g$.
\label{fig-2}}
\end{figure}

\begin{figure}
\figurenum{3}
\epsscale{1.00}
\plotfiddle{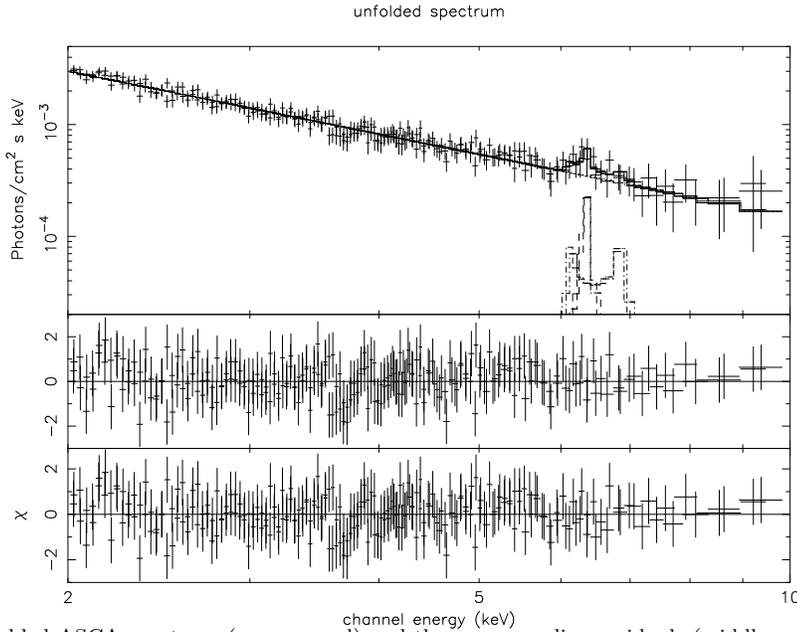}{6.cm}{270}{45}{45}{-200}{250}
\caption[Lu_figure3.ps]{
 The unfolded ASCA spectrum (upper panel) and the corresponding
residuals (middle panel) for the best fitted truncated disk model with an
inclination angle 20 degree. For comparison, the residuals for a power-law
plus a Gaussian line fit are shown in the bottom panel. Only SIS data
between 2-10 keV are shown for the clarity.
\label{fig-3}}
\end{figure}

\end{document}